\documentclass[
prd,twocolumn,showpacs,preprintnumbers,nofootinbib,amsmath,amssymb,floatfix]{revtex4-1}
\usepackage{graphicx}
\usepackage{epstopdf}
\usepackage{amsmath,amssymb}
\newcommand{\Mpl}{M_{\textrm{pl}}}
\newcommand{\Ms}{M_{\ast}}
\newcommand{\Mmes}{M_{\textrm{mess}}}
\newcommand{\mG}{m_{\tilde{G}}}
\newcommand{\mg}{m_{\tilde{g}}}
\newcommand{\mga}{m_{\tilde{g},1}}
\newcommand{\mgb}{m_{\tilde{g},2}}
\newcommand{\mgc}{m_{\tilde{g},3}}

\newcommand{\mNLSP}{m_{\textrm{NLSP}}}
\newcommand{\TeV}{\textrm{TeV}}
\newcommand{\GeV}{\textrm{GeV}}
\newcommand{\MeV}{\textrm{MeV}}
\newcommand{\keV}{\textrm{keV}}
\newcommand{\eV}{\textrm{eV}}

\newcommand{\meter}{\textrm{m}}

\newcommand{\cm}{\textrm{cm}}

\newcommand{\sgn}{\textrm{sgn}}
\newcommand{\cgrav}{c_{\textrm{grav}}}

\newcommand{\ddecay}{d_{\textrm{decay}}}

\newcommand{\mweak}{m_{\textrm{weak}}}
\newcommand{\mSUSY}{m_{\textrm{SUSY}}}
\newcommand{\ipb}{\textrm{pb}^{-1}}
\newcommand{\ifb}{\textrm{fb}^{-1}}
\renewcommand{\eqref}[1]{Eq.~(\ref{#1})}

\newcommand{\secref}[1]{Sec.~\ref{sec:#1}}

\newcommand{\gravitino}{\tilde{G}}

\begin{document}

\preprint{UCI-TR-2010-05}

\title{Light Gravitinos at Colliders and Implications for Cosmology}

\author{Jonathan L. Feng$^1$, Marc Kamionkowski$^2$, and Samuel K. Lee$^2$}
\affiliation{$^1$Department of Physics and Astronomy, University of California, Irvine, CA 92697\\ 
$^2$California Institute of Technology, Mail Code 350-17, Pasadena, CA 91125}

\date{\today}

\begin{abstract}
Light gravitinos, with mass in the eV to MeV range, are well-motivated
in particle physics, but their status as dark-matter candidates is
muddled by early-Universe uncertainties.  We investigate how upcoming
data from colliders may clarify this picture.  Light gravitinos are
produced primarily in the decays of the next-to-lightest
supersymmetric particle, resulting in spectacular signals, including
di-photons, delayed and non-pointing photons, kinked charged tracks,
and heavy metastable charged particles.  We find that the Tevatron
with $20~\ifb$ and the 7 TeV LHC with $1~\ifb$ may both see evidence
for hundreds of light-gravitino events.  Remarkably, this collider
data is also well suited to distinguish between currently viable
light-gravitino scenarios, with striking implications for structure
formation, inflation, and other early-Universe cosmology.
\end{abstract}

\pacs{14.80.Ly, 13.85.-t, 95.35.+d, 98.80.Cq}
\maketitle

\section{Introduction} \label{sec:intro}

Supersymmetry is one of the most promising ideas for new physics
beyond the Standard Model.  Supersymmetric theories that incorporate
local supersymmetry (or supergravity) predict the existence of the
gravitino, the spin-3/2 superpartner of the graviton.  When
supersymmetry is broken, the gravitino acquires a mass through the
super-Higgs mechanism, ``eating'' the spin-1/2 goldstino, the
Goldstone fermion associated with spontaneously-broken local
supersymmetry \cite{Volkov:1973jd,Fayet:1974jb,de Wit:1975th,%
Deser:1977uq}.  In contrast to other superpartners, the gravitino can
have a mass $\mG$ that is not at the weak scale ${\mweak \sim 100~\GeV
- 1~\TeV}$, and viable models exist for gravitino masses as low as the
eV scale and as high as 100 TeV.  In this work, we consider light
gravitinos, with mass in the eV to MeV range.  Such gravitinos are
highly motivated in particle physics, as they emerge in models with
gauge-mediated supersymmetry breaking (GMSB), in which constraints on
flavor violation are naturally satisfied~\cite{Dine:1981za,%
Dimopoulos:1981au,Nappi:1982hm,AlvarezGaume:1981wy,Dine:1994vc,%
Dine:1995ag}.

Light gravitinos also have cosmological motivations.  In particular,
they are the original supersymmetric dark-matter
candidate~\cite{Pagels:1981ke}.  Assuming a high reheating
temperature, gravitinos are initially in thermal equilibrium and then
freeze out while still relativistic.  As we discuss in detail below,
their resulting relic density is
\begin{equation}
\Omega_{\tilde{G}} h^2 \simeq \left[ \frac{\mG}{1~\keV} \right] 
\left[ \frac{106.75}{g_{*S,f}} \right] \ ,
\label{omega}
\end{equation}
where $g_{*S,f}$ is the number of relativistic degrees of freedom at
freeze out, and has been normalized to the total number of degrees of
freedom in the Standard Model.  When originally proposed in the
1980's, uncertainties in $h$ and the total matter relic density
allowed $\mG \sim \keV$.  This led to a simple and attractive
gravitino--dark-matter scenario, consistent with standard Big Bang
cosmology, in which the Universe cooled from some high temperature,
and keV gravitinos froze out and now form all of the dark matter.

In the intervening years, however, a variety of astrophysical
constraints have greatly complicated this picture.  First, the
dark-matter relic density is now known to be ${\Omega_{\textrm{DM}}
h^2 \simeq 0.11}$.  Second, constraints on structure formation, as
probed by galaxy surveys and Lyman-$\alpha$ forest observations,
require that the bulk of dark matter be cold or
warm~\cite{Viel:2005qj}.  As we will discuss more fully below, this
leads to three scenarios of interest:
\begin{enumerate}
\setlength{\itemsep}{1pt}\setlength{\parskip}{0pt}\setlength{\parsep}{0pt}
\item $\mG \alt 15-30~\eV$: Gravitinos are produced by the standard
cosmology leading to \eqref{omega}; they are hot dark matter, but
their contribution is small enough to be consistent with the observed
small-scale structure.  Some other dark-matter particle is required.
\item $15-30~\eV \alt \mG \alt \textrm{few}~\keV$: Non-standard cosmology
and a non-standard gravitino production mechanism are required, both
to avoid overclosure and to cool the gravitinos to satisfy
small-scale-structure constraints.  Some other dark-matter particle
may be required.
\item $\mG \agt \textrm{few}~\keV$: Non-standard cosmology is required
to dilute the thermal relic density of \eqref{omega}.  Gravitinos
produced by thermal freeze out are cold enough to be all of the dark
matter.
\end{enumerate}
Note that the original ``keV gravitino'' scenario, previously favored,
is now the most disfavored, in the sense that it is excluded by both
overclosure and small-scale-structure constraints.  All of the
possibilities are rather complicated, however, as in each case, some
additional physics is required, either to provide the rest of the dark
matter or to modify the history of the early Universe to allow
gravitinos to be all of the dark matter.

In this paper, we discuss how collider data may help clarify this
picture.  Light gravitinos are primarily produced at colliders in the
decays of the next-to-lightest supersymmetric particle (NLSP).  It is
a remarkable coincidence that modern particle detectors, with
components placed between 1 cm to 10 m from the beamline, are
beautifully suited to distinguish between the NLSP decay lengths
predicted in scenarios 1, 2, and 3.  For example, the decay length of
a Bino NLSP decaying to a gravitino is~\cite{Cabibbo:1981er}
\begin{equation}
c \tau \simeq 23~\cm \left[ \frac{\mG}{100~\eV} \right]^2
\left[ \frac{100~\GeV}{m_{\tilde{B}}} \right]^5 \ .
\end{equation}
This implies that scenarios 1, 2, and 3 make distinct predictions for
collider phenomenology, and the identification of the gravitino
collider signatures realized in nature may have far-reaching
implications for the early Universe.

Of course, this requires that gravitinos can be produced in sufficient
numbers and distinguished from Standard Model backgrounds.  In this
work, we determine event rates for a variety of signatures, including
prompt di-photons and delayed and non-pointing photons (relevant for
neutralino-NLSP scenarios), as well as kinked charged tracks and heavy
metastable charged particles (relevant for stau-NLSP scenarios).  We
present results for an assumed final Tevatron dataset ($20~\ifb$ of
$2~\TeV$ $p\bar p$ collisions), an early LHC dataset ($1~\ifb$ of
$7~\TeV$ $pp$ collisions), and a future LHC dataset ($10~\ifb$ of
$14~\TeV$ $pp$ collisions).  We find that the final Tevatron and early
LHC data have roughly equivalent sensitivity to these events, with
both capable of seeing hundreds of distinctive light gravitino events.
The full LHC data greatly extends the reach in parameter space, and
may also allow precision measurements of NLSP lifetimes and gravitino
masses.

We begin in \secref{cosmology} by reviewing the cosmological bounds on
light gravitinos and discussing how these bounds are relaxed in
early-Universe scenarios that differ from the canonical one.  In
\secref{colliders} we then discuss NLSP decays to gravitinos, GMSB
models, and current collider constraints.  In \secref{prospects} we
present our results for the number of light-gravitino events at
colliders, based on collider simulations, and discuss the cosmological
implications.  We summarize our conclusions in \secref{conclusion}.

\section{Light Gravitino Cosmology}
\label{sec:cosmology}

\subsection{Canonical Scenario}

\subsubsection{Relic Abundance}

In the currently canonical scenario, after inflation, the Universe is
reheated to a temperature $T_R$ that is assumed to be far higher
(e.g., $10^{12}$ or $10^{15}~\GeV$) than the weak scale.  During this
phase, inelastic scattering processes and decays can convert
Standard-Model particles in the thermal bath into
gravitinos~\cite{Weinberg:1982zq,Ellis:1984eq,Moroi:1993mb,%
Kawasaki:1994af,Borgani:1996ag}.  The rate $C_{\tilde{G}}$ per unit
volume for production of light gravitinos (strictly speaking, only the
spin-1/2 goldstino components) can be calculated by considering all
such processes, which primarily involve strong~\cite{Bolz:2000fu} and
electroweak gauge bosons~\cite{Pradler:2006qh,Pradler:2007ne}, as well
as top quarks~\cite{Rychkov:2007uq}.  The total result, valid in the
limit ${T \gg \mSUSY}$, where $\mSUSY$ is the scale of the
superpartner masses, is~\cite{Rychkov:2007uq}
\begin{equation} \label{eqn:CG}
    C_{\tilde{G}} \simeq 15 \frac{\mg^2}{\mG^2} \frac{T^6}{\Mpl^2} \ ,
\end{equation}
where ${\Mpl \simeq 1.2 \times 10^{19}~\GeV}$ is the Planck mass.
Here we have assumed that the gaugino masses $\mga$, $\mgb$, and
$\mgc$ and the tri-linear scalar coupling $A_t$ are at a common mass
scale.  For simplicity, we have set them equal to a universal gaugino
mass $\mg$.

The evolution of the gravitino number density $n_{\tilde G}$ via these
production processes, and their inverses, is governed by the Boltzmann
equation
\begin{equation}
    \frac{dn_{\tilde G}}{dt} +  3 H n_{\tilde G} = C_{\tilde G}
    - \Gamma n_{\tilde G} \ ,
\label{eqn:boltzmann}
\end{equation}
where $H$ is the Hubble expansion rate and $\Gamma$ is the rate of
processes that annihilate gravitinos.  The $3H n_{\tilde G}$ term
accounts for dilution of the number density due to cosmological
expansion.  If $\Gamma \gg H$, gravitinos are in thermal equilibrium,
${\Gamma n_{\tilde G} = C_{\tilde G}}$, and their number density (the
solution to the Boltzmann equation) is
\begin{equation}
    n_{\tilde G}^{\mathrm{eq}} = g \frac{2 \zeta(3)}{\pi^2}T^3
    \simeq 0.24\, T^3 \ .
\label{eqn:gravitinoabundance}
\end{equation}
Here we used ${g=2}$, since it is primarily the spin-1/2 goldstino
components that are produced thermally.

The rate $\Gamma$ at which a given gravitino is destroyed in the
plasma is then
\begin{equation}
    \Gamma = \frac{C_{\tilde G}}{n_{\tilde G}^{\mathrm{eq}}} \simeq 60\,
    \frac{ m_{\tilde g}^2 T^3}{m_{\tilde G}^2 \Mpl^2} \ .
\label{eqn:Gamma}
\end{equation}
Since ${\Gamma \propto T^3}$ and ${H \propto T^2}$, the ratio
${\Gamma/H \propto T}$ is largest at the highest temperatures.  Thus,
if ${\Gamma(T_R) \agt H(T_R)}$ at reheating, then gravitinos come into
thermal equilibrium shortly after reheating.  During this era, the
expansion rate is given by ${H \simeq 1.66\, g_*^{1/2} T^2/\Mpl}$;
assuming reheating temperatures ${T_R \gg \TeV}$, at which all
particles in the minimal supersymmetric Standard Model (MSSM) are
relativistic, we set the number $g_*$ of relativistic degrees of
freedom to ${g_*(T_R)\simeq 228.75}$.  Comparing $\Gamma(T_R)$ and
$H(T_R)$, we then see that if the reheating temperature satisfies
\begin{equation}
    T_R \agt  T_f \equiv 5~\GeV \left[ \frac{\mG}{\keV} \right]^2
\left[ \frac{\TeV}{m_{\tilde g}} \right]^2 \ ,
\label{eqn:Treheat}
\end{equation}
then gravitinos come into thermal equilibrium after reheating.
Recalling that the production rate, Eq.~(\ref{eqn:CG}), used here is
valid only for ${T \gg \mNLSP}$ (i.e., ${T \agt 10~\TeV}$), we
conclude that for weak-scale gluino masses, light gravitinos with
${\mG \alt \MeV}$ will come into thermal equilibrium if the reheating
temperature is ${T_R\agt 5\times10^6~\GeV}$.

The creation/annihilation rates for gravitinos at temperatures ${T
\alt 10~\TeV}$ have not yet been calculated, and so the precise
temperature at which gravitinos freeze out (which occurs when ${\Gamma
\simeq H}$) cannot yet be determined.  Still, gravitinos are produced
and destroyed individually, requiring (from R-parity conservation)
that each creation/destruction is accompanied by creation/destruction
of some other supersymmetric particle.  Therefore, the freezeout
temperature $T_f$ cannot be much lower than the mass $\mNLSP$ of the
NLSP, as the equilibrium abundance of SUSY particles then decreases
exponentially.  We thus conclude that the freezeout temperature for
light gravitinos falls roughly in the range ${10~\GeV \alt T_f \alt
10~\TeV}$.

With this range of freezeout temperatures, ${T_f \gg \mG}$, so
gravitinos are relativistic when they freeze out.  The relic gravitino
density is then~\cite{Pagels:1981ke}
\begin{equation}
    \Omega_{\tilde G} h^2 \simeq 0.1 \left[ \frac{m_{\tilde
    G}}{100~\eV} \right]
    \left[\frac{106.75}{g_{*S,f}}\right] \ ,
\label{eqn:thermalabundance}
\end{equation}
the standard result for hot relics with ${g=2}$, where $g_{*S,f}$ is
the number of relativistic degrees of freedom when the gravitinos
freeze out.  If gravitinos freeze out when all of the MSSM degrees of
freedom are relativistic, ${T_f \gg m_{\mathrm{SUSY}}}$, and
${g_{*S,f}=228.75}$.  However, it is more likely that freeze out
occurs at ${T\sim \mNLSP \sim 100~\GeV}$, when $g_{*S,f} \sim 100$
\cite{Pierpaoli:1997im}.

\subsubsection{Cosmological Constraints}

Given that current cosmic-microwave-background (CMB) and
structure-formation measurements constrain the density of dark matter
to be ${\Omega_{\textrm{DM}} h^2 \simeq 0.11}$,
Eq.~(\ref{eqn:thermalabundance}) implies an upper bound ${\mG \alt
200~\eV}$.  The upper limit ${\mG \simeq 200~\eV}$ is saturated if the
gravitino makes up {\em all} of the dark matter and freezeout occurs
when ${g_{*S,f}=228.75}$ is the maximum value allowed in the MSSM.
However, a gravitino of this mass would be hot dark matter.  It would
smooth density perturbations on scales probed by galaxy surveys and
the Lyman-$\alpha$ forest to a degree that is highly inconsistent with
data.  As $\mG$ is reduced from this upper limit, the smoothing scale
is increased (the gravitinos get ``hotter''), but the gravitino
abundance is reduced, thus making the magnitude of the smoothing
smaller.  A combination of data from the cosmic microwave background,
galaxy surveys, and the Lyman-$\alpha$ forest constrain the
contribution of a hot component of dark matter to be
$\alt15\%$~\cite{Viel:2005qj}, implying for ${g_{*S,f} \simeq 100}$
that ${\mG \alt 15~\eV}$.  This suggests that the most conservative
upper bound is given by ${\mG \alt 30~\eV}$, in the case that
${g_{*S,f} \simeq 200}$ approaches the maximal value allowed in the
MSSM.  Therefore, in this canonical scenario, thermal gravitinos with
mass less than $30~\eV$ make up only a fraction of the dark matter,
thus requiring some other particle to be the cold dark matter.  This
is the first cosmological scenario listed in Sec.~\ref{sec:intro}.

We conclude by noting that future astrophysical data are likely to
improve.  And while the current sensitivity is to gravitino masses as
small as ${15-30~\eV}$, it is forecast that next-generation
experiments may be sensitive to gravitino masses as small as ${1~\eV}$
\cite{Ichikawa:2009ir}.  A collider detection of a gravitino in the
mass range ${\mG=1-30~\eV}$ would thus lead to testable consequences
in forthcoming cosmological data.

\subsection{Non-standard Early-Universe Scenarios}

There are several ways in which the early-Universe production of relic
gravitinos could differ from the canonical scenario outlined above.
Thus, there are scenarios in which a gravitino of mass ${\mG \agt
30~\eV}$, ruled out in the canonical model, could be cosmologically
consistent or, better yet, completely compose the dark matter.

Let us first consider scenarios in which the gravitinos reach thermal
equilibrium in the early Universe, since most observational
constraints are strictly valid only under this assumption.  As
mentioned above, if we only consider particles in the MSSM, then
${g_{*S,f} \leq 228.75}$, and Eq.~(\ref{eqn:thermalabundance})
suggests an upper limit of ${\mG \alt 200~\eV}$ from the relic
abundance constraint.  One way to evade this limit is to simply
consider higher values of $g_{*S,f}$; i.e., gravitinos decouple and
freeze out earlier than in the canonical scenario.  This may be
possible in models with more degrees of freedom than the MSSM.  More
massive gravitinos that decouple earlier may then be viable, if they
have an abundance that obeys the constraint ${\Omega_{\textrm{DM}} h^2
\alt 0.11}$.

It is possible that this constraint is saturated and that these
heavier gravitinos entirely compose the dark matter.  Of course, we
must still require that these heavier gravitinos are not so hot as to
erase structure to a degree that contradicts observations.  The same
combination of CMB, galaxy-survey, and Lyman-$\alpha$--forest data
that was used to constrain ${\mG \alt 30~\eV}$ in the canonical
scenario can also be used to constrain the gravitino mass in this
early-decoupling scenario, assuming that thermal gravitinos make up
all of the dark matter.  With this assumption, Ref.~\cite{Viel:2005qj}
find ${\mG \agt 550~\eV}$, using a selection of Lyman-$\alpha$ data.
The same authors later find a stronger constraint, ${\mG \agt
2~\keV}$, with SDSS Lyman-$\alpha$ data
\cite{Viel:2006kd,Boyarsky:2008xj}, a result slightly weaker than a
bound on warm-dark-matter models obtained by
Ref.~\cite{Seljak:2006qw}.  A number of other small-scale observations
also seem to support that ${\mG \agt \textrm{few}~\keV}$ under these
assumptions \cite{Primack:2009jr}.

We thus conclude that if ${30~\eV \alt \mG \alt \textrm{few}~\keV}$,
then thermal gravitinos are too warm to be the only component of the
dark matter, regardless of whether or not they have the correct
abundance.  Gravitinos in this mass range would only be viable if some
other non-standard early-Universe process cools them, or if there is
an additional cold component.  This is the second scenario mentioned
in Sec.~\ref{sec:intro}.  However, if ${\mG \agt \textrm{few}~\keV}$,
then gravitinos may be sufficiently cold, and may in early-decoupling
scenarios have the right abundance, to be the dark matter.  This is
the third scenario outlined in Sec.~\ref{sec:intro}.

Of course, aside from early decoupling, there are other non-standard
mechanisms that can reduce the gravitino abundance.  For example,
recall that we have no empirical constraints to the early Universe
prior to the epoch of big-bang nucleosynthesis (BBN), at which $T \agt
\textrm{few}~\MeV$ \cite{Hannestad:2004px}. Thus, some
entropy-producing process prior to BBN could also dilute the gravitino
abundance.  It is possible that there may be some exotic
early-Universe physics that conspires to produce the same effect.  One
relatively simple possibility is that the reheating temperature is
low.  If the reheating temperature is smaller than the freezeout
temperature, then gravitinos will never come into thermal equilibrium,
and their relic abundance will thus be accordingly
smaller~\cite{Asaka:2000zh,Allahverdi:2004si,Copeland:2005qe,%
Kohri:2005wn,Steffen:2006hw,Pradler:2006hh,Choi:2007rh}.  The only
catch is that for the light gravitinos we consider here, the reheating
temperature must be unusually low for this to occur.  For example, if
${\mG=\keV}$ and ${\mg=300~\GeV}$, then Eq.~(\ref{eqn:Treheat})
suggests that the reheating temperature must be ${T_R\alt 50~\GeV}$.
However, recall that this estimate may not be strictly valid at ${T
\alt 10~\TeV}$, as we have already noted.  Thus, a more careful
calculation of the production rate of light gravitinos at low
reheating temperatures may be necessary.  Nevertheless, such low
reheating temperatures have been considered \cite{Kohri:2009ka}, and
Ref.~\cite{Gorbunov:2008ui} has examined an explicit low-reheat
scenario in which a gravitino of mass ${\mG=1-15~\keV}$ can have the
right abundance to be the dark matter.

Finally, we also note that there may be additional mechanisms
affecting the generation of gravitinos.  For example, in our
discussion we have neglected the nonthermal contribution to the
gravitino abundance from out-of-equilibrium decays of other
supersymmetric particles.  There may also be other significant modes
of gravitino production or dilution, including processes involving the
messenger particles responsible for GMSB~\cite{Dimopoulos:1996gy,%
Choi:1999xm,Baltz:2001rq,Fujii:2002fv,Jedamzik:2005ir,Staub:2009ww},
nonthermal production via oscillations of the inflaton
field~\cite{Kallosh:1999jj}, and various other
mechanisms~\cite{deGouvea:1997tn}.  There may thus be other reasons
why the gravitino abundance or temperature differs from those in the
canonical thermal-production scenario; this may be true even if ${\mG
\alt 30~\eV}$.

To summarize, in the canonical model, gravitinos are required to have
mass ${\mG\alt 30~\eV}$ and form only a fraction of the dark matter.
Gravitinos with mass range ${\mG \agt 30~\eV}$ would require
non-standard physics or cosmology to reduce their abundance or
temperature to agree with observations.  Below we discuss collider
signatures of light gravitinos.  We close here by noting that such
collider data may, if gravitinos are discovered, thus help
discriminate between the diversity of early-Universe scenarios for
gravitino production.

\section{Light Gravitinos at Colliders} 
\label{sec:colliders}

\subsection{Mass and Interactions}

The gravitino mass is determined by the super-Higgs mechanism.  In
simple models, it is given in terms of the supersymmetry-breaking
scale $F$, which has mass dimension 2, as
\begin{equation}
\mG = \frac{F}{\sqrt{3} \Ms} \simeq
240~\eV \left[ \frac{\sqrt{F}}{10^3~\TeV} \right]^2 \ ,
\label{mass}
\end{equation}
where ${\Ms \equiv \Mpl/\sqrt{8\pi} \simeq 2.4 \times 10^{18}~\GeV}$
is the reduced Planck mass.

The interactions of weak-scale gravitinos are of gravitational
strength, as expected since they are the superpartners of gravitons.
However, the couplings of the goldstino are proportional to $1/F$
\cite{Casalbuoni:1988kv,Lee:1998aw}.  The interactions of light
gravitinos are therefore dominated by their goldstino components, and
may be much stronger than gravitational.  Decays to gravitinos are
faster for light gravitinos.

For reasons to be discussed below, we will focus on cases where the
NLSP is either the neutralino or the stau.  For a neutralino NLSP that
is dominantly a Bino, the decay widths to gravitinos
are~\cite{Cabibbo:1981er,Ambrosanio:1996jn}
\begin{eqnarray}
\Gamma(\tilde{B} \to \gamma \gravitino) &=&
\frac{\cos^2 \theta_W m_{\tilde{B}}^5}{16 \pi F^2} \\
\Gamma(\tilde{B} \to Z \gravitino) &=&
\frac{\sin^2 \theta_W m_{\tilde{B}}^5}{16 \pi F^2} 
\left[ 1 - \frac{m_Z^2}{m_{\tilde{B}}^2} \right]^4 \ ,
\end{eqnarray}
where $\theta_W$ is the weak mixing angle.  For ${m_{\tilde{B}} \alt
m_Z}$, decays to $Z$ bosons are negligible or kinematically forbidden,
and the corresponding decay length is
\begin{equation}
c \tau \simeq 23~\cm \left[ \frac{\mG}{100~\eV} \right]^2
\left[ \frac{100~\GeV}{m_{\tilde{B}}} \right]^5 \ .
\end{equation}
For heavier neutralinos, the $Z$ mode may be significant; for very
heavy Binos, the branching ratio for this mode is ${B(Z) \simeq \sin^2
\theta_W \simeq 0.23}$.  Decays to ${l^+l^-\gravitino}$, where
${l=(e,\mu,\textrm{or}\,\tau)}$ is a charged lepton, and $h\gravitino$
may also be possible; however, these modes have branching ratios of
$\sim 0.01$ and $\sim 10^{-6}$, respectively.

For stau NLSPs, the decay width is~\cite{Feng:1997zr}
\begin{equation}
\Gamma(\tilde{\tau} \to \tau \gravitino) =
\frac{m_{\tilde{\tau}}^5}{16 \pi F^2} \ ,
\end{equation}
corresponding to a decay length
\begin{equation}
c \tau \simeq 18~\cm \left[ \frac{\mG}{100~\eV} \right]^2
\left[ \frac{100~\GeV}{m_{\tilde{\tau}}} \right]^5 \ .
\end{equation}

As anticipated, in both the neutralino-NLSP and stau-NLSP scenarios,
the decay lengths for gravitinos in the cosmologically interesting
range correspond to distances that bracket the size of collider
detectors.

\subsection{GMSB Models}

Light gravitinos are expected to be dominantly produced at colliders
in the cascade decays of strongly-interacting superpartners, such as
squarks and gluinos.  Collider constraints therefore depend on the
full superpartner spectrum, and so are model-dependent.  Following
most of the literature, we will work in the framework of minimal GMSB,
and so we briefly review its features here.

Typical GMSB models are characterized by a hidden sector, a messenger
sector, and a visible sector, the MSSM.  Supersymmetry breaking is
triggered by a hidden-sector gauge-singlet superfield $S$ acquiring
the vacuum expectation value ${S = M + \theta^2 F_S}$.  This then
generates masses for the messenger-sector fields $\Mmes = \lambda M$,
where $\lambda$ is a coupling in the superpotential.  These in turn
generate masses for the visible-sector superpartners that are roughly
a loop factor times ${\Lambda \equiv F_S/M}$, and so ${\Lambda \sim
100~\TeV}$.  Note that ${\Mmes > \Lambda}$ is generally assumed.

In the minimal GMSB framework, the entire superpartner spectrum is
specified by the parameters
\begin{equation}
\Lambda,\ \Mmes,\ N_5,\ \tan{\beta},\ \sgn(\mu),\ \cgrav \ .
\label{parameters}
\end{equation}
Here, $\Lambda$ and $\Mmes$ are as described above; masses and
couplings are generated at $\Mmes$ and then evolved to the weak scale
via the renormalization group.  The number of messenger superfields is
given by $N_5$, the effective number of $\mathbf{5} +
\mathbf{\bar{5}}$ representations of SU(5).  The Higgs sector is
specified by the usual parameters $\tan\beta$ and $\sgn(\mu)$.  The
last parameter is
\begin{equation}
\cgrav \equiv \frac{F}{\lambda F_S} \ ,
\end{equation}
where $F = (F_S^2 + \sum_i F_i^2 )^{1/2}$ is the total
supersymmetry-breaking vacuum expectation value, which appears in
\eqref{mass}.  These relations imply
\begin{equation} \label{eqn:mG}
\mG = \cgrav \frac{\Mmes \Lambda}{\sqrt{3} \Ms} \ .
\end{equation}
We expect ${\cgrav \agt 1}$, since ${F \ge F_S}$ and ${\lambda \alt
1}$, and in the minimal case that there is only one non-zero $F$-term,
we expect ${\cgrav \sim 1}$.

The superpartner masses are determined by the parameters of
\eqref{parameters}; for details, see Ref.~\cite{Baer:2006rs}.  Here we
note only two things.  First, the superpartner masses are determined
by gauge couplings.  Thus, although, for example,
chargino~\cite{Kribs:2008hq} and sneutrino~\cite{Santoso:2009qa} NLSPs
have been considered, the canonical NLSP candidates are those with
only hypercharge interactions, namely, the Bino and right-handed
sleptons.  Among the right-handed sleptons, the stau is typically the
lightest, as renormalization-group evolution and left-right--mixing
effects both decrease the stau mass relative to the selectron and
smuon, and so we will focus on the Bino-NLSP and stau-NLSP
scenarios.\footnote{Note that in the ``slepton co-NLSP'' scenario,
where the three charged sleptons are degenerate to within the mass of
the tau, the number of $\tilde{e} \to e\gravitino$ and $\tilde{\mu}
\to \mu\gravitino$ decays may be comparable to that of the
$\tilde{\tau} \to \tau\gravitino$ decay that usually dominates
gravitino production.}  Second, the Bino and stau masses are
proportional to $N_5$ and $\sqrt{N_5}$, respectively.  For ${N_5=1}$,
the NLSP is the Bino in minimal GMSB, but for ${N_5 > 1}$, the stau
may also be the NLSP; see, for example, Fig.~1 of
Ref.~\cite{Feng:1997zr}.

Thus, to study the Bino-like neutralino-NLSP scenario, we will choose
${N_5=1}$; likewise, we choose ${N_5=4}$ to study the stau-NLSP
scenario.  For both scenarios, we fix ${\tan \beta = 20}$, ${\mu>0}$,
and ${\cgrav=1}$.  We let $\Lambda$ and $\Mmes$ be free parameters.
Note that the overall mass scale of the supersymmetric partners is
roughly proportional to $\Lambda$, while the gravitino mass depends on
both $\Lambda$ and $\Mmes$ as in Eq.~(\ref{eqn:mG}).  Thus, scanning
over the free GMSB parameters will allow us to explore collider
signals for a range of masses.  We shall now examine the existing
collider constraints on the parameter spaces of these two scenarios.

\subsection{Current Collider Constraints}

The high-energy collider signals of GMSB and gravitinos are well
studied~\cite{Stump:1996wd,Dimopoulos:1996vz,Dimopoulos:1996va,%
Ambrosanio:1996jn,Feng:1997zr,Abreu:2000nm,Baer:2000pe,%
Ambrosanio:2000ik,Pagliarone:2003ya,Hamaguchi:2004df,Feng:2004yi,%
Wagner:2004bx,Martyn:2006as,Hamaguchi:2007ji,Tarem:2009zz,Chen:2009gu};
for a review of current bounds, see Ref.~\cite{Feng:2009te}.  Here we
summarize the most relevant results for the models and signals we
consider below.

We shall discuss GMSB signals in more detail below, but we summarize
them briefly here.  In the neutralino-NLSP scenario, there are several
possible signals.  For short-lived neutralinos, all supersymmetry
events include two prompt high energy photons.  For longer-lived
neutralinos that travel a macroscopic distance before decaying to
photons in the detector, delayed or non-prompt photons are possible.
The stau-NLSP scenario may also lead to a variety of signatures,
depending on the stau lifetime, including acoplanar leptons, tracks
with large impact parameters, kinked charged tracks, and heavy
metastable charged particles.

Several studies have attempted to place constraints on GMSB models by
searching for these signals.  Given that we will scan over a large
range of the GMSB parameter space, we are primarily interested in
constraints that are generally valid over this entire range.  We shall
thus focus on limits from LEP studies, based on an integrated
luminosity of ${628~\ipb}$ at center-of-mass energies of
${189-209~\GeV}$, which combined
searches for both GMSB and neutral-Higgs
signals~\cite{Heister:2002vh,lepgmsb}.  The relevant results for our
models are the lower limits of ${\Lambda \agt 70~\TeV}$ for our
neutralino-NLSP model, and ${\Lambda \agt 20~\TeV}$ for our stau-NLSP
model; see Fig.~6 of Ref.~\cite{Heister:2002vh}.  These constraints on
$\Lambda$ are valid for all values of $\Mmes$ we include in our scan.
Therefore, the allowed region of ${\Mmes-\Lambda}$ parameter space is
constrained by these LEP bounds.

However, there are also a number of studies that focused on
constraining specific benchmark models~\cite{Allanach:2002nj}, which
occupy certain points or lines in the GMSB parameter space.  Although
these constraints cannot be directly applied to our models, we discuss
them to get an idea of the robustness of the LEP bounds on our
parameter space.

Of these benchmark-model constraints, the best collider bounds on
di-photon events are from the Tevatron, including a D0 search based on
an integrated luminosity of $1.1~\ifb$~\cite{:2007is} and a CDF search
based on $2.6~\ifb$~\cite{Aaltonen:2009tp}.  The D0 and CDF bounds,
when interpreted assuming the benchmark GMSB model SPS 8 from
Ref.~\cite{Allanach:2002nj}, lead to lower bounds on the Bino mass of
$125~\GeV$ and $150~\GeV$, respectively.  For longer-lived neutralinos
that travel a macroscopic distance before decaying to photons in the
detector, a CDF search for delayed photons, based on $570~\ipb$ of
data, established lower bounds on $m_{\tilde{B}}$ from 70 to 100 GeV
for neutralino decay lengths between 20 cm and 6 m, again when
interpreted in the context of SPS 8~\cite{Abulencia:2007ut}.

Searches for heavy metastable charged particles have also been
performed at D0~\cite{Abazov:2008qu}, assuming the benchmark GMSB
model SPS 7 from Ref.~\cite{Allanach:2002nj}.  A similar search was
performed at CDF~\cite{Aaltonen:2009kea}, but did not interpret
results in the context of GMSB models.  Based on $\sim 1~\ifb$ of
data, and assuming only Drell-Yan slepton production, the constraints
resulting from these two searches are not competitive with the LEP
bounds stated previously.

Thus, we shall take the more general LEP bounds as constraints on the
two models we consider in this work, and shall further take only
conservative values of the lower limits.  For the neutralino NLSP
model, we shall only scan the parameter space with ${\Lambda \geq
80~\TeV}$, which should be comfortably allowed by the LEP bounds.
However, we acknowledge that it is possible that the Tevatron data may
exclude a small range of NLSP masses within this parameter space
comparable to that ruled out in the benchmark model (i.e., $\alt
150~\GeV$), should this data be reanalyzed in the context of our
models.  For the stau NLSP model, we shall scan over ${\Lambda \geq
30~\TeV}$.  Given that the current Tevatron constraints are not
competitive with the LEP bounds, all of this parameter space should be
allowed.  As we will see, hadron colliders have bright prospects for
probing the parameter spaces of these models.

\section{Tevatron and LHC Prospects}
\label{sec:prospects}

\subsection{Gravitino Signals}

The collider signal of a supersymmetric particle decaying to a
gravitino can be classified by (1) the distance from the interaction
point at which the decay occurs, and (2) the nature of the
accompanying Standard Model decay products.  The former is determined
by the gravitino mass and the masses of the decaying supersymmetric
particles, as well as the speed with which the decaying particles are
produced.  The latter is determined primarily by the nature of the
NLSP.  We shall define and investigate the following categories of
events:
\begin{enumerate}
\setlength{\itemsep}{1pt}\setlength{\parskip}{0pt}\setlength{\parsep}{0pt}

\item Prompt di-photons (in neutralino NLSP models): Events in which
two photons are produced (via a pair of neutralino decays to
gravitinos) within $d_{\textrm{pr}}$ of the interaction point.  We
take ${d_{\textrm{pr}} = 1~\cm}$ as a conservative estimate of the
distance to which the origin of any photon can be resolved in
detectors at the Tevatron and the LHC.  Note that here and below, we
cut on the total distance traveled by the NLSP before it decays, not
its (transverse) distance from the beamline when it decays.

\item Non-prompt photons (neutralino NLSP): Events in which at least
one photon is produced at a mid-detector distance $\ddecay$ away from
the interaction point, where ${d_{\textrm{pr}} \leq \ddecay \leq
d_{\textrm{np}}}$, and $d_{\textrm{np}}$ is the maximum distance from
the interaction point at which a photon can be observed.  We
conservatively take $d_{\textrm{np}} = 3~\meter$, roughly the outer
radius of the hadronic calorimeters at both the Tevatron and the LHC.
(Note that although the calorimeters in the ATLAS detector at the LHC
actually extend to $\sim4~\meter$, those in the CMS detector only
extend to $\sim3~\meter$; we have thus taken the more conservative
$3~\meter$ as our cut.)  Photons may also convert and be seen in the
muon chambers, extending the sensitivity to decays $\sim
10~\meter$ from the interaction point, but we neglect this possibility
here.  Here we also take ${d_{\textrm{pr}} =
1~\cm}$.

\hspace{1.5em} Note that this category of events encompasses both
non-pointing photons and delayed photons.  A non-pointing photon is
simply a photon that does not spatially point back to the interaction
point.  A delayed photon has the further distinction of being produced
only after a significant temporal delay following the time of the
initial collision.  This may occur when the particle that decays to
the photon is produced with a low speed, so that it takes a
non-negligible amount of time to travel away from the interaction
point before it decays.  If this amount of time is comparable to the
time between collision events, it may be difficult to properly
identify the delayed photon with its originating event.

\item Non-prompt leptons (stau NLSP): Events in which at least one
charged lepton is produced (via charged-slepton decays to gravitinos)
at a mid-detector distance $\ddecay$, where ${d_{\textrm{pr}} \leq
\ddecay \leq d_{\textrm{np}}}$ as before.  We take
${d_{\textrm{np}}=5~\meter}$ and ${d_{\textrm{np}}=7~\meter}$ as the
outer radii of the muon chambers in the detectors at the Tevatron and
the LHC, respectively.  (As above, although the muon chambers in the
ATLAS detector at the LHC extend to $\sim10~\meter$, those in the CMS
detector only extend to $\sim7~\meter$; we take the more conservative
$7~\meter$ as our cut.)  We again take ${d_{\textrm{pr}} = 1~\cm}$.
Each of these events produces a distinctive charged track with a kink
due to the momentum carried away by the gravitino.

\hspace{1.5em} As above, both non-pointing and delayed events are
included in this category.  Furthermore, we include all generations
($e$, $\mu$, and $\tau$).  As mentioned previously, the stau is
generally the lightest slepton, and hence we expect the majority of
the decays in the stau-NLSP scenario to be of the form ${\tilde{\tau}
\to \tau\gravitino}$.  Although the heavier sleptons
${\tilde{l}=(\tilde{e}\,\textrm{or}\,\tilde{\mu})}$ may also decay to
$l\gravitino$, the branching ratio of this decay is generally
suppressed compared to the decay to a lepton and a neutralino, i.e.,
$l\tilde{\chi}^0$.  If the latter is kinematically forbidden, then the
3-body decays to $l\tau^-\tilde{\tau}^+$ or $l\tau^+\tilde{\tau}^-$
dominate instead.  However, as the mass splitting between the stau
NLSP and the heavier sleptons decreases, these 3-body decays become
less dominant (becoming kinematically forbidden if the mass splitting
becomes less than the tau mass).  The decays to $l\gravitino$ may then
occur if the heavier sleptons $\tilde{l}$ are produced at the end of a
decay chain.

\item Metastable sleptons (stau NLSP): Events in which at least one
charged slepton passes through the entire detector before decaying to
a charged lepton and a gravitino.  That is, the gravitino is produced
at ${\ddecay \geq d_{\textrm{ms}}}$, where $d_{\textrm{ms}}$ is the
distance to the outer edge of the detector.  We take ${
d_{\textrm{ms}}=5~\meter}$ and ${ d_{\textrm{ms}}=10~\meter}$ as
conservative estimates of the sizes of the detectors at the Tevatron
and the LHC, respectively.  All generations ($\tilde{e}$,
$\tilde{\mu}$, and $\tilde{\tau}$) are included.  These events will
produce charged tracks with a relatively large radius of curvature.

\hspace{1.5em} For this category, we impose a further cut, requiring
that the speeds $\beta$ of the sleptons satisfy the criteria
${\beta_{\textrm{lower}} \leq \beta \leq \beta_{\textrm{upper}}}$.
The lower cut removes slower sleptons, which may be identified with
the incorrect collision event.
The higher cut removes faster sleptons, which may be misidentified as
muons.  We take typical values
${\beta_{\textrm{lower}}=0.6}$ and ${\beta_{\textrm{upper}}=0.8}$.
Note, however, that Ref.~\cite{Chen:2009gu} suggests a new search
strategy that may be sensitive to even higher values of $\beta$.
\end{enumerate}

All of these events will also be distinguished by missing energy and
momentum carried away by the gravitinos.  Note that these categories
are chosen to be illustrative of the variety of signals that may be
observed, and that they are not comprehensive---we do not investigate
prompt di-lepton events or neutralino decays to $Z$ bosons, for example.
Furthermore, the categories are not mutually exclusive; for example,
one may easily have a single event in which both a non-prompt lepton
and a metastable slepton are produced.  It is also clear that the
relevant detector systematics and backgrounds will also be different
for each category.

This categorization of events is somewhat oversimplified, as it is
based primarily on cuts on the decay length.  Certainly, additional
cuts will be required in a realistic analysis, possibly reducing the
number of detected signals.  However, we shall soon see that these
simple categories align with the three cosmological scenarios outlined
previously.

\subsection{GMSB Scan and Collider Simulations}

We now calculate the event rates for these gravitino signals in a
parametrized GMSB model.  A large number of programs have been written
for the numerical computation of the mass spectra and collider
predictions for parametrized supersymmetric models
\cite{Skands:2003cj,Allanach:2008qq}.  In this paper, we use ISAJET
7.80/ISASUSY ~\cite{Paige:2003mg} to generate mass spectra and decay
branching ratio tables.  ISASUSY properly includes a number of 3-body
decay processes relevant for gravitino phenomenology that are missing
in other branching ratio programs.

ISAJET/ISASUSY takes values of the GMSB parameters listed in
Eq.~(\ref{parameters}) as input.  As discussed previously, here we
focus on parameterizations that fix a subset of the GMSB parameters,
resulting in either a neutralino or a stau NLSP.  We then scan over
$\Mmes$ and $\Lambda$ (requiring that ${\Mmes > \Lambda}$), resulting
in spectra with a range of gravitino and NLSP masses.  The
correspondence between the ${\Mmes-\Lambda}$ scan and the resulting
${\mG-\mNLSP}$ parameter space is shown in Fig.~\ref{fig:refplots}.

\begin{figure*}
$\begin{array}{cc}
\includegraphics[width=.75\columnwidth]{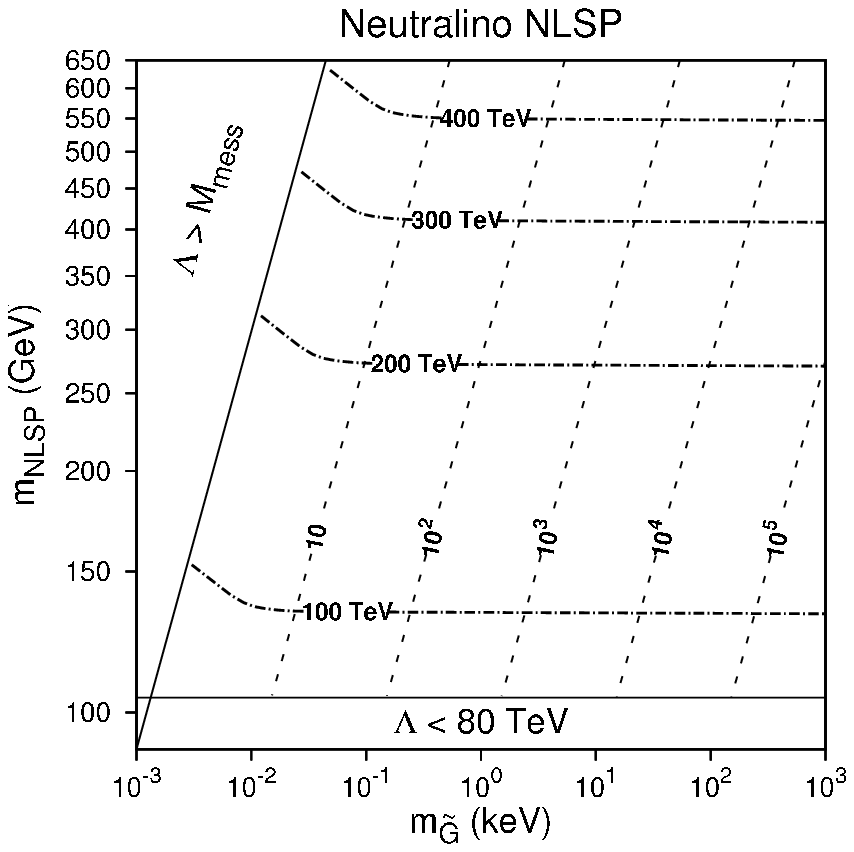} &
\includegraphics[height=.75\columnwidth]{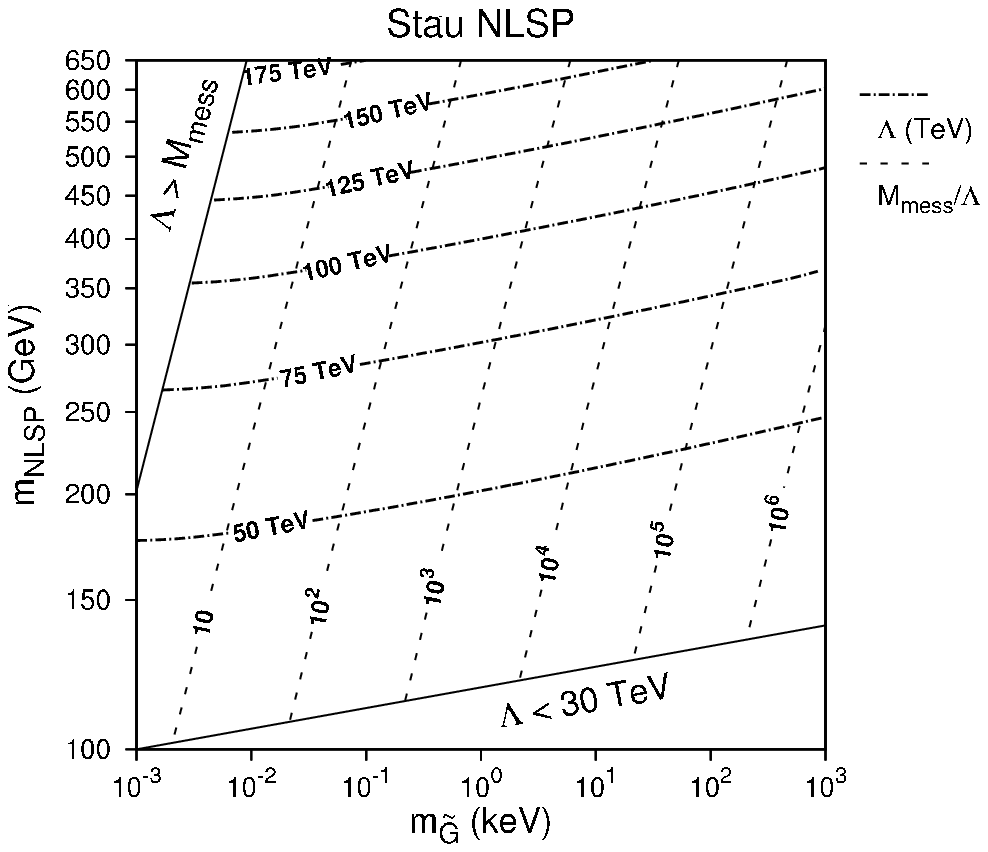}
\end{array}$
\caption{Plots showing the mapping between the ${\mG-\mNLSP}$ and the
${\Mmes-\Lambda}$ GMSB parameter spaces, for the neutralino-NLSP
scenario with $N_5 = 1$ (left) and stau-NLSP scenario with $N_5 = 4$
(right), where we fix $\tan \beta = 20$, $\mu > 0$, and $\cgrav = 1$
in both cases.  Contours of constant ${\Mmes/\Lambda}$ (dashed black)
and $\Lambda$ (dash-dotted black) are shown.  The region in the
upper-left corner is disallowed by theory, while the region at the
bottom is excluded by experiment (using the conservative constraints
mentioned in the text).  }
\label{fig:refplots}
\end{figure*}

We then take the spectra and decay tables output by ISAJET/ISASUSY and
use them as input for the Monte Carlo event generator PYTHIA
6.4.22~\cite{Sjostrand:2006za}, including all supersymmetric processes
available therein.  For a given center-of-mass energy, PYTHIA can
simulate a given number of collision events, giving a complete record
of the various decay chains and final products generated in each event
and an estimation of the various production cross sections.  From this
record, we can identify the supersymmetric ``mother'' particles that
decay to directly produce gravitino and Standard Model ``daughter''
particles in each individual event.  We can also find the decay length
$\ddecay$ away from the interaction point that each mother particle
travels before decaying to produce a gravitino.  Thus, for any number
of simulated events, we can find the fraction that fall into each of
the above categories.  The expected number of signals from each
category is then given by the respective fraction multiplied by the
total number of supersymmetric events.  We can also calculate the
average $\langle\ddecay\rangle$ of the decay length, taken over all
supersymmetric events.

\subsection{Cosmological Implications}

The results of the scan are shown in
Figs.~\ref{fig:neutp}-\ref{fig:staums}.  We can see that the simple
categorization of collider signals by decay-length cuts corresponds
surprisingly well with the categorization of cosmological scenarios
outlined previously.  For example, Fig.~\ref{fig:neutp} shows that the
observation of hundreds of prompt events suggests that the first
cosmological scenario (${\mG \alt 30~\eV}$) is likely to be valid.
Likewise, the second cosmological scenario (${30~\eV \alt \mG \alt
\textrm{few}~\keV}$) will be implied by the observation of a large
number of non-prompt events, as demonstrated by
Figs.~\ref{fig:neutnp}~and~\ref{fig:staunp}.  Finally, that the
observation of a large number of metastable sleptons supports the
third cosmological scenario (${\mG \agt \textrm{few}~\keV}$) can be
seen in Fig.~\ref{fig:staums}.  We emphasize that this correspondence
is not strongly dependent on our specific choice of GMSB models.  It
is indeed a remarkable coincidence that theoretically-motivated
supersymmetric and gravitino mass scales, the physical sizes of
collider detectors, and gravitino cosmology all conspire to allow this
correspondence.

Note also that we find that the number of gravitino events produced
during the initial run of the LHC (center-of-mass energy of $7~\TeV$
and integrated luminosity of $1~\ifb$) may be comparable to that
produced during an extended run of the Tevatron (center-of-mass energy
of $2~\TeV$ and integrated luminosity of $20~\ifb$).  This is true in
regions of parameter space where large numbers of signals are
expected.  However, the higher center-of-mass energy of the LHC allows
it to access regions of parameter space where $\mSUSY$ is larger; this
is especially evident in the neutralino-NLSP scenario, as can be seen
by comparing the left and middle panels in both
Figs.~\ref{fig:neutp}~and~\ref{fig:neutnp}.

If the distribution of mother-particle decay lengths can be measured
with sufficient accuracy along with the total signal rate, then it may
be possible to gain some information on the masses of the mother
particles and the gravitino.  To do so, it will be important to
understand the distribution of energies and speeds with which mother
particles are produced, since this will directly affect the
distribution of decay lengths via dilation of the mother-particle
lifetimes.  In Fig.~\ref{fig:betagamma}, we show some examples of
probability distribution functions for the speed $\beta$ and the
Lorentz factor $\gamma$ of mother particles that decay to gravitinos,
for various collider scenarios.

\begin{figure*}
$\begin{array}{ccc}
\includegraphics[width=.6\columnwidth]{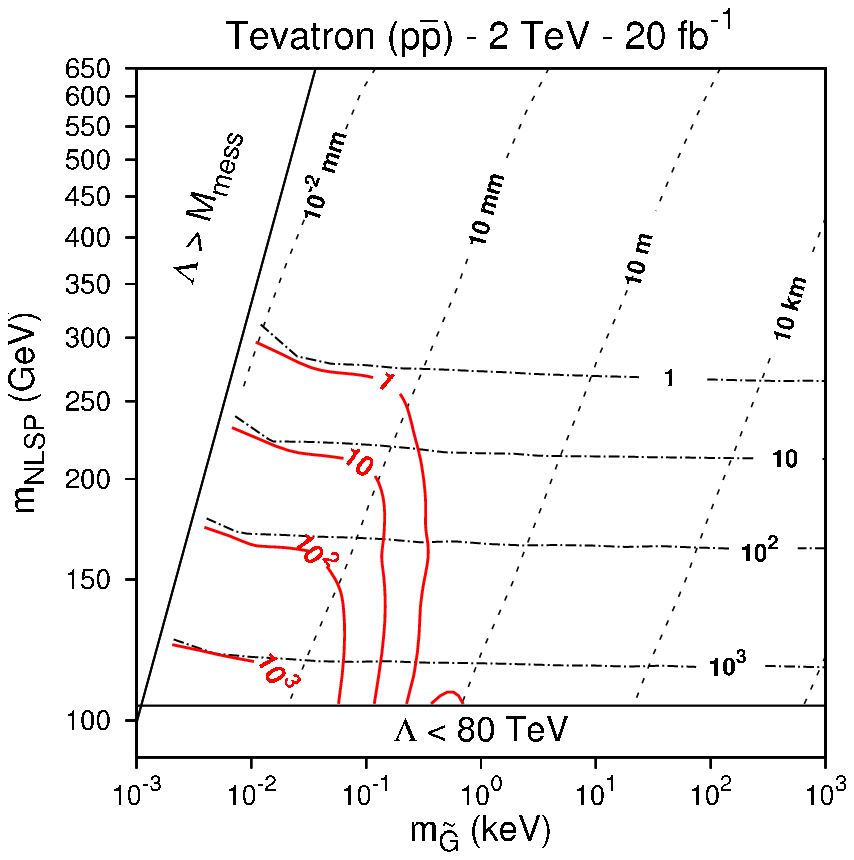} &
\includegraphics[width=.6\columnwidth]{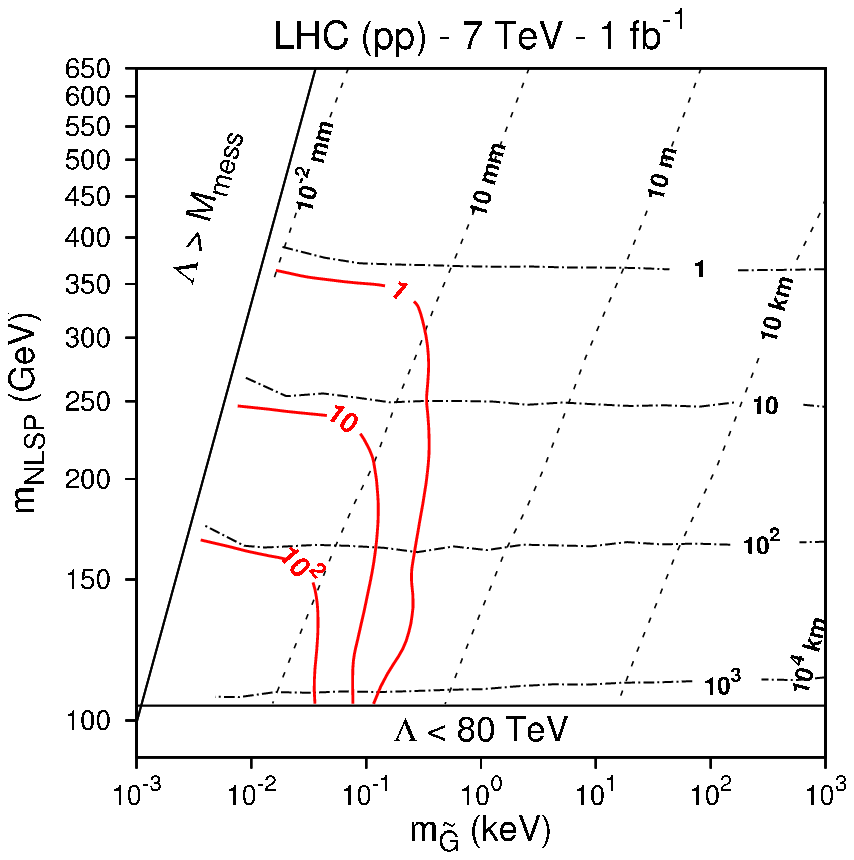} &
\includegraphics[height=.6\columnwidth]{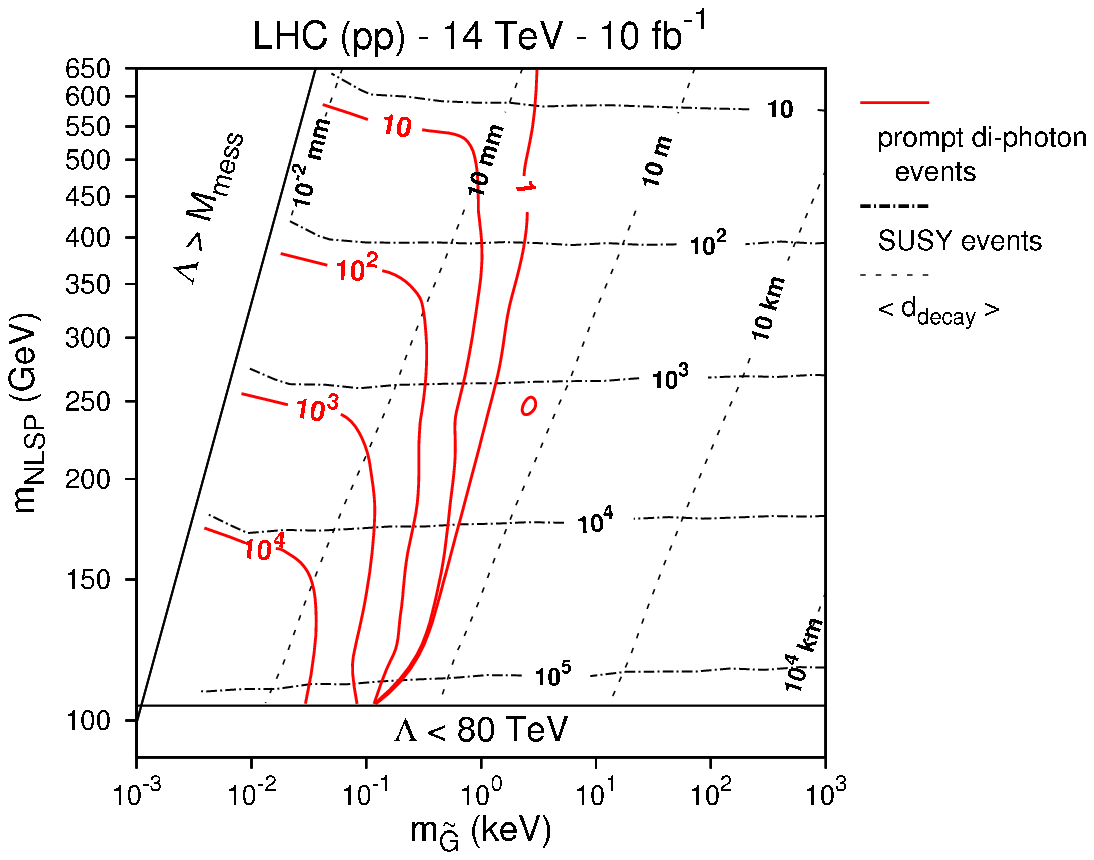}
\end{array}$
\caption{Contour plots over the ${\mG-\mNLSP}$ parameter space showing
the expected number of prompt di-photon events (solid red) in a model
with a neutralino NLSP, for the three collider scenarios (indicated at
the top of each plot) of interest.  The total number of supersymmetric
events (dash-dotted black) and the average decay length $\ddecay$
(dashed black) expected at each point in the parameter space are also
indicated by contours.  The region in the upper-left corner is
disallowed if we require ${\Lambda < \Mmes}$, while the region at the
bottom is ruled out by LEP (using the conservative constraints
discussed in the text).  Note that hundreds of signal events may occur
at the Tevatron with $20~\ifb$ and at the $7~\TeV$ LHC with $1~\ifb$
if ${\mG \alt \textrm{tens of}~\eV}$.  Observation of such a number of
events would suggest that the canonical thermal-production scenario is
correct, and that light gravitinos compose only a fraction of the dark
matter.  Also, note that a larger fraction of neutralinos instead
decay to $Z\gravitino$ as the neutralino mass increases (and that
there is some small fraction of decays to $e^+e^-\gravitino$ and
$h\gravitino$).}
\label{fig:neutp}
\end{figure*}

\begin{figure*}
$\begin{array}{ccc}
\includegraphics[width=.6\columnwidth]{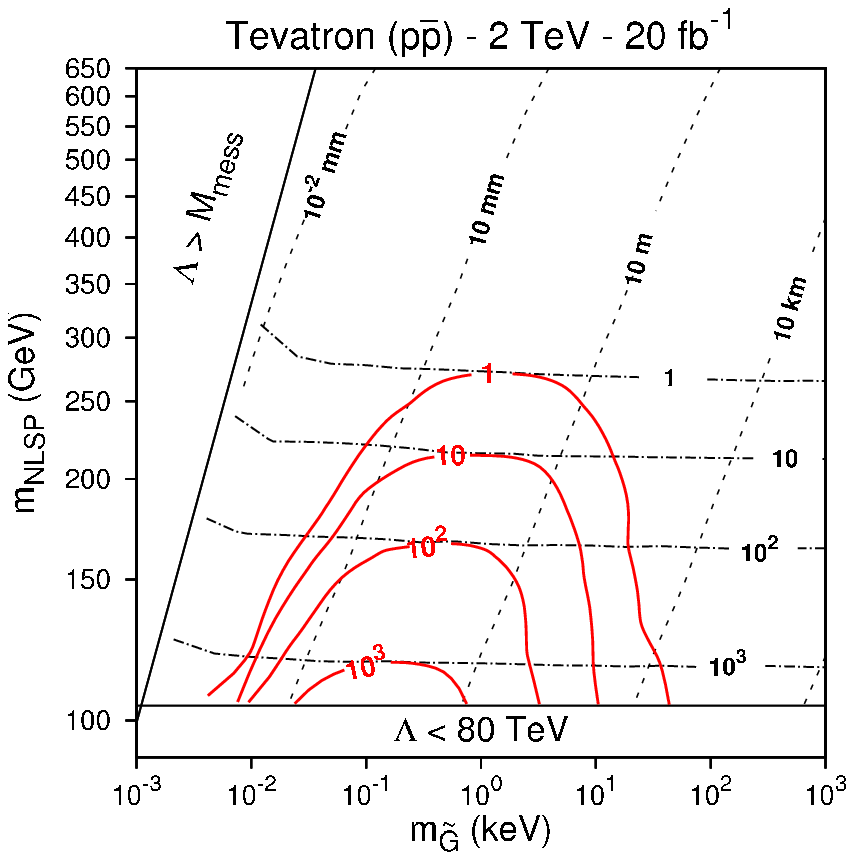} &
\includegraphics[width=.6\columnwidth]{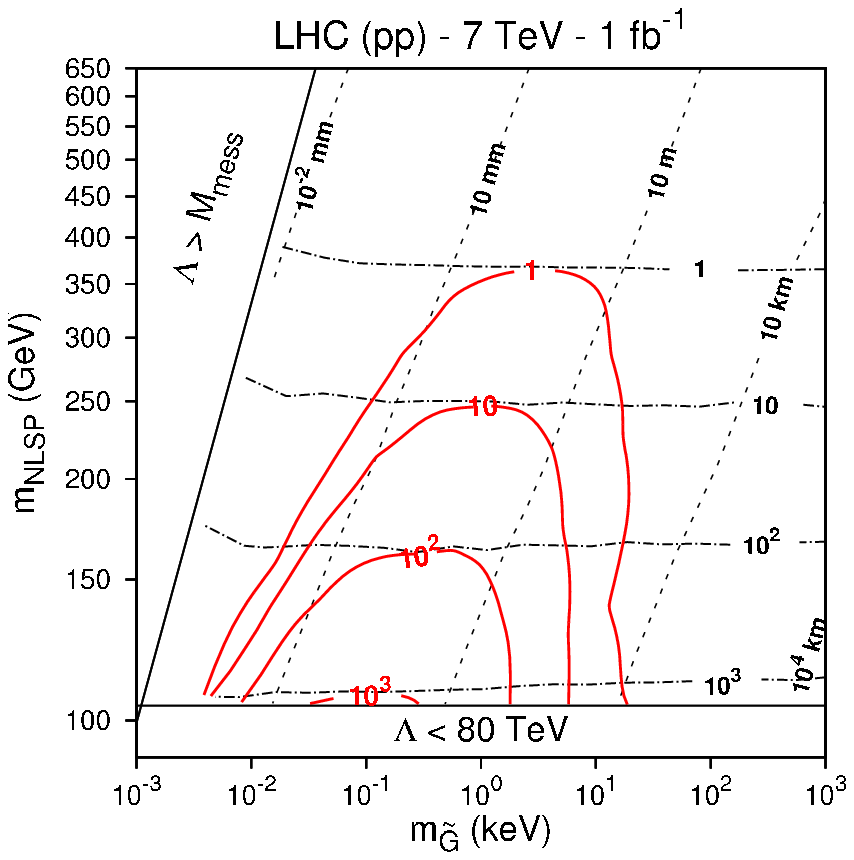} &
\includegraphics[height=.6\columnwidth]{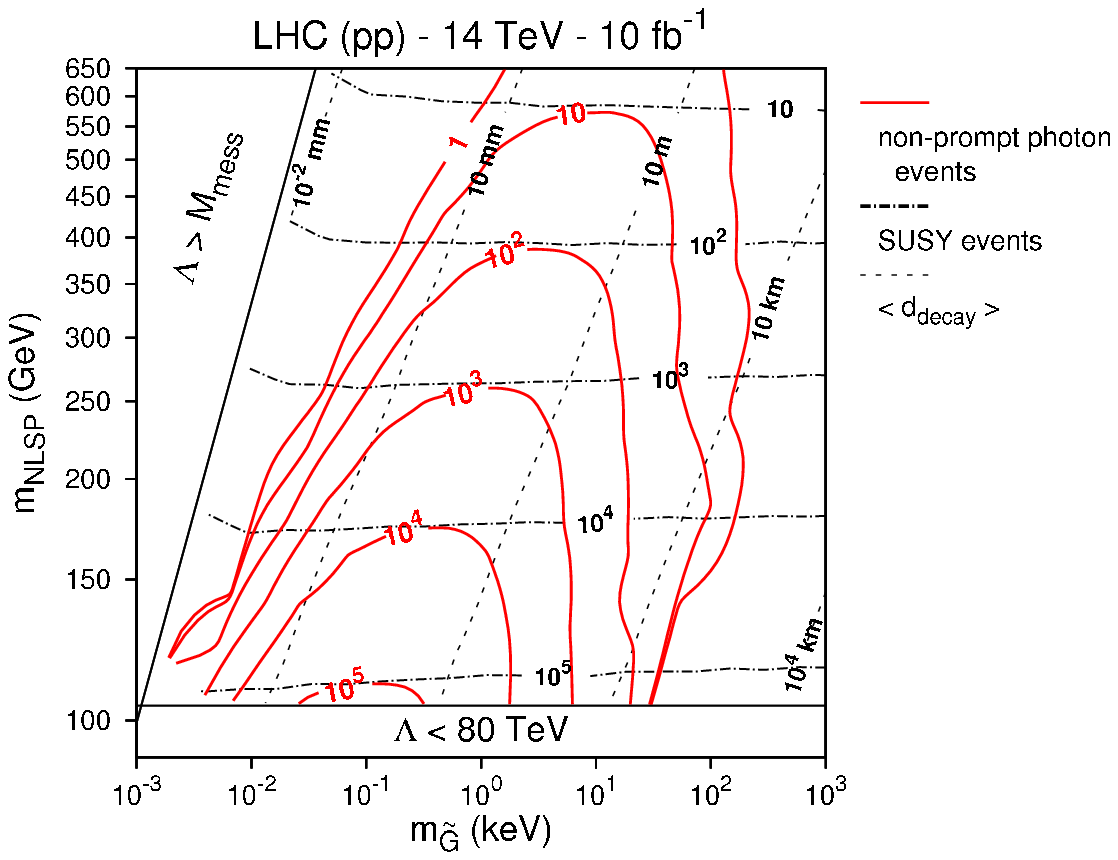}
\end{array}$
\caption{The same as Fig.~\ref{fig:neutp}, but for non-prompt photon
events.  Note that hundreds of signal events may occur at the Tevatron
with $20~\ifb$ and at the $7~\TeV$ LHC with $1~\ifb$ if ${\textrm{tens
of}~\eV \alt \mG \alt \textrm{few}~\keV}$.  Observation of this number
of events would suggest that a non-standard cosmology and gravitino
thermal history cooled relic gravitinos, and, for the top part of this
mass range, also diluted the relic density.}
\label{fig:neutnp}
\end{figure*}

\begin{figure*}
$\begin{array}{ccc}
\includegraphics[width=.6\columnwidth]{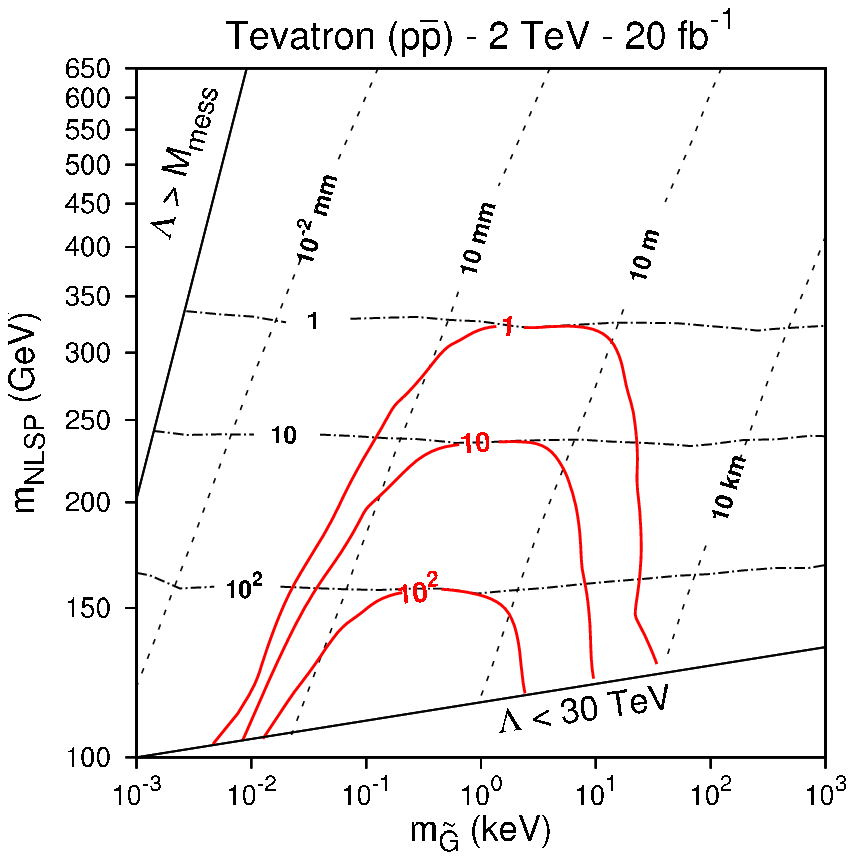} &
\includegraphics[width=.6\columnwidth]{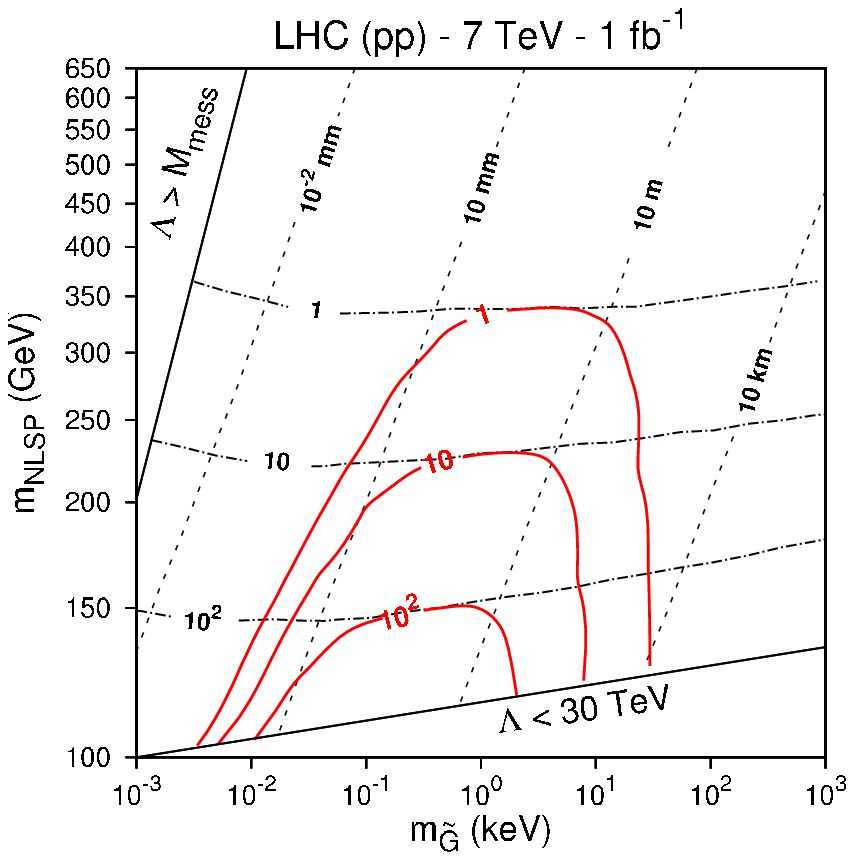} &
\includegraphics[height=.6\columnwidth]{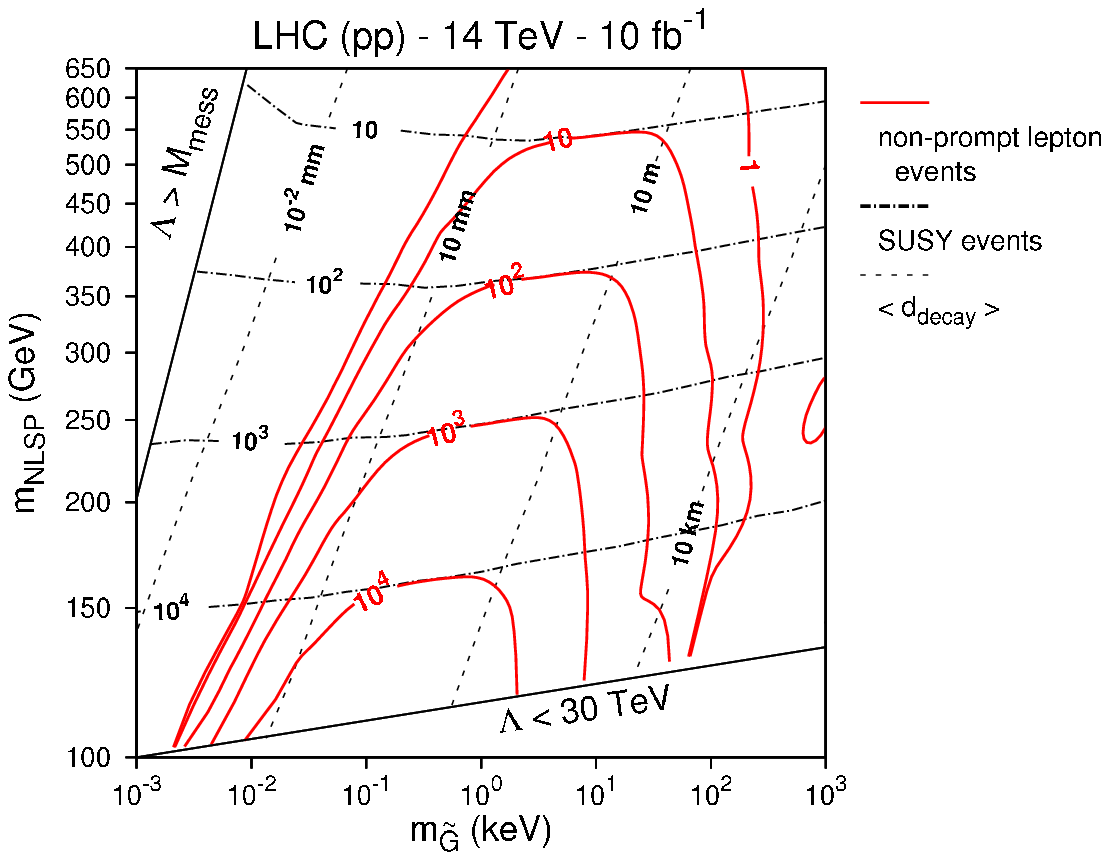}
\end{array}$
\caption{Contour plots showing the expected number of non-prompt
lepton events in a model with a stau NLSP.  Note that hundreds of
signal events may occur at the Tevatron with $20~\ifb$ and at the
$7~\TeV$ LHC with $1~\ifb$ if ${\textrm{tens of}~\eV \alt \mG \alt
\textrm{few}~\keV}$.  Observation of this number of events would
suggest that a non-standard cosmology and gravitino thermal history
cooled relic gravitinos, and, for the top part of this mass range, also
diluted the relic density.}
\label{fig:staunp}
\end{figure*}

\begin{figure*}
$\begin{array}{ccc}
\includegraphics[width=.6\columnwidth]{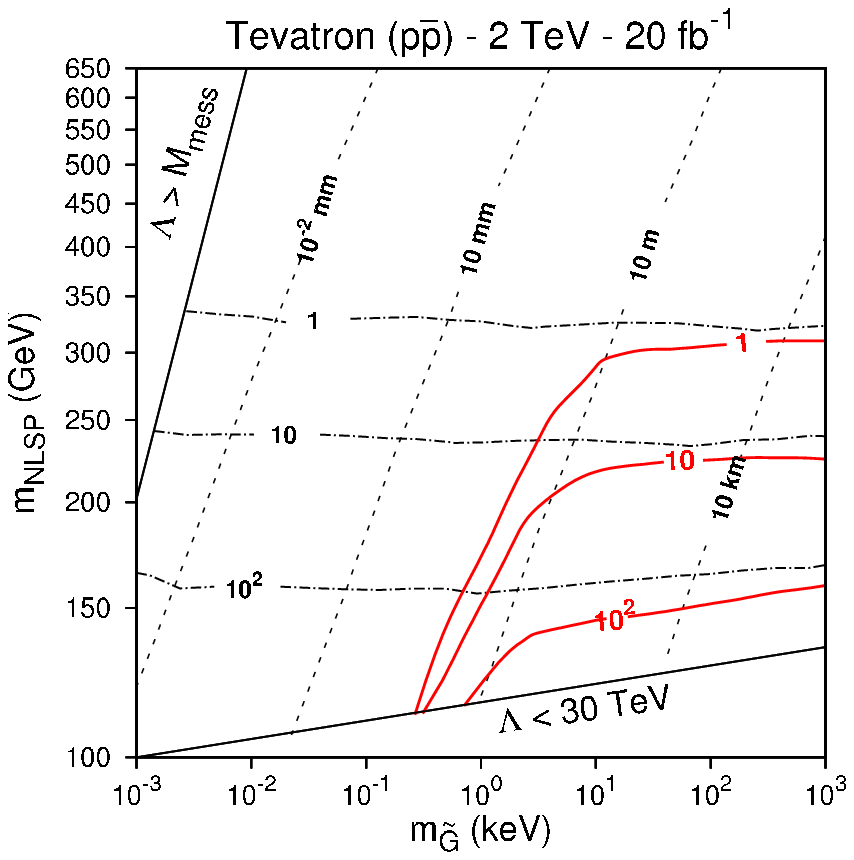} &
\includegraphics[width=.6\columnwidth]{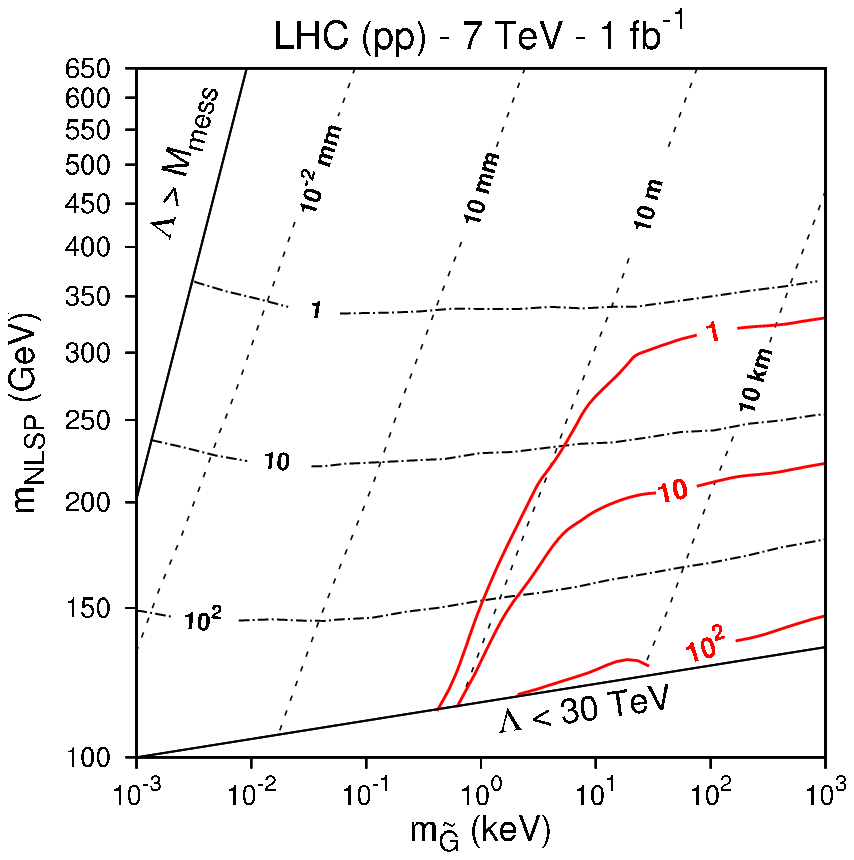} &
\includegraphics[height=.6\columnwidth]{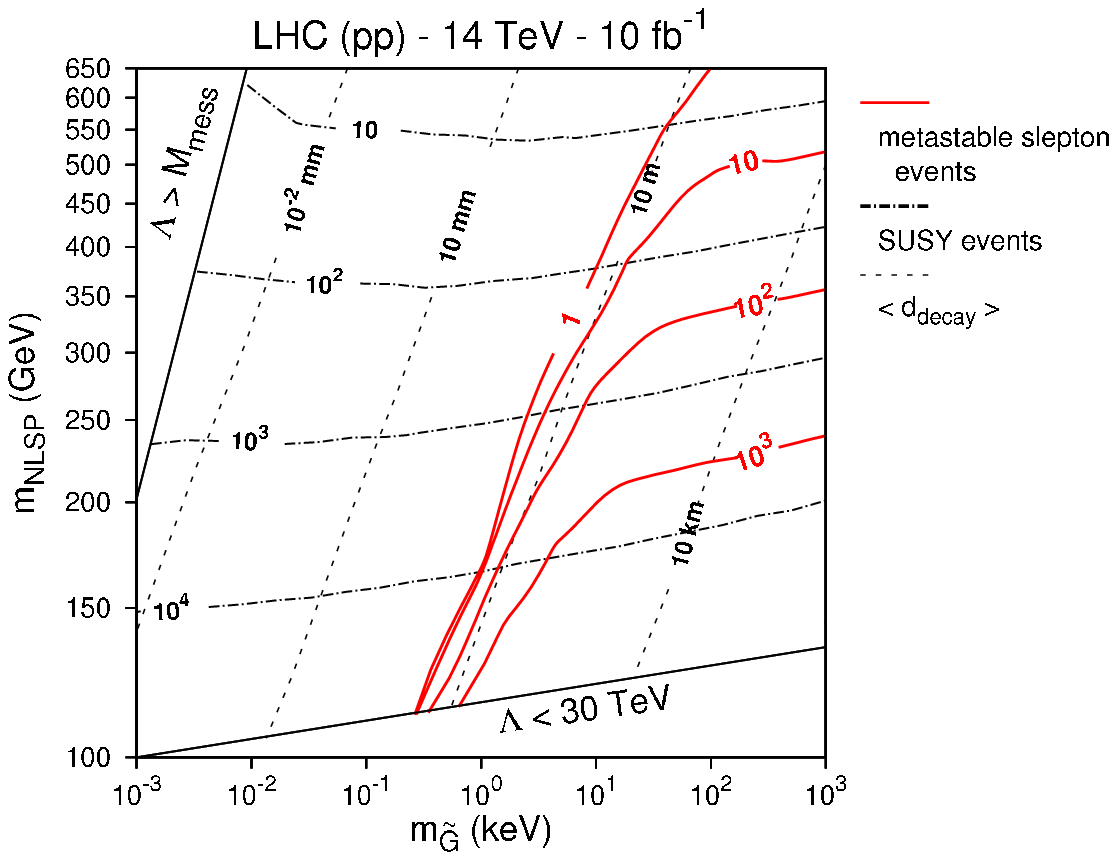}
\end{array}$
\caption{The same as Fig.~\ref{fig:staunp}, but for metastable slepton
events.  Note that hundreds of signal events may occur at the Tevatron
with $20~\ifb$ and at the $7~\TeV$ LHC with $1~\ifb$ if ${\mG \agt
\keV}$.  Observation of such a number of events would suggest that
gravitinos could entirely compose the dark matter, assuming some
non-standard cosmology diluted relic gravitinos.  Also, note that
although we only consider ${\mG \alt \MeV}$ here, GMSB models allow
larger gravitino masses, up to ${\mG \sim \GeV}$.  Thus, these plots
may be straightforwardly extrapolated to higher gravitino masses if
desired.  However, note that at higher $\Mmes$, and hence at higher
$\mG$, the neutralino again becomes the NLSP; see Fig.~1 of
Ref.~\cite{Feng:1997zr}.}
\label{fig:staums}
\end{figure*}

\begin{figure*}
$\begin{array}{cc}
\includegraphics[width=\columnwidth]{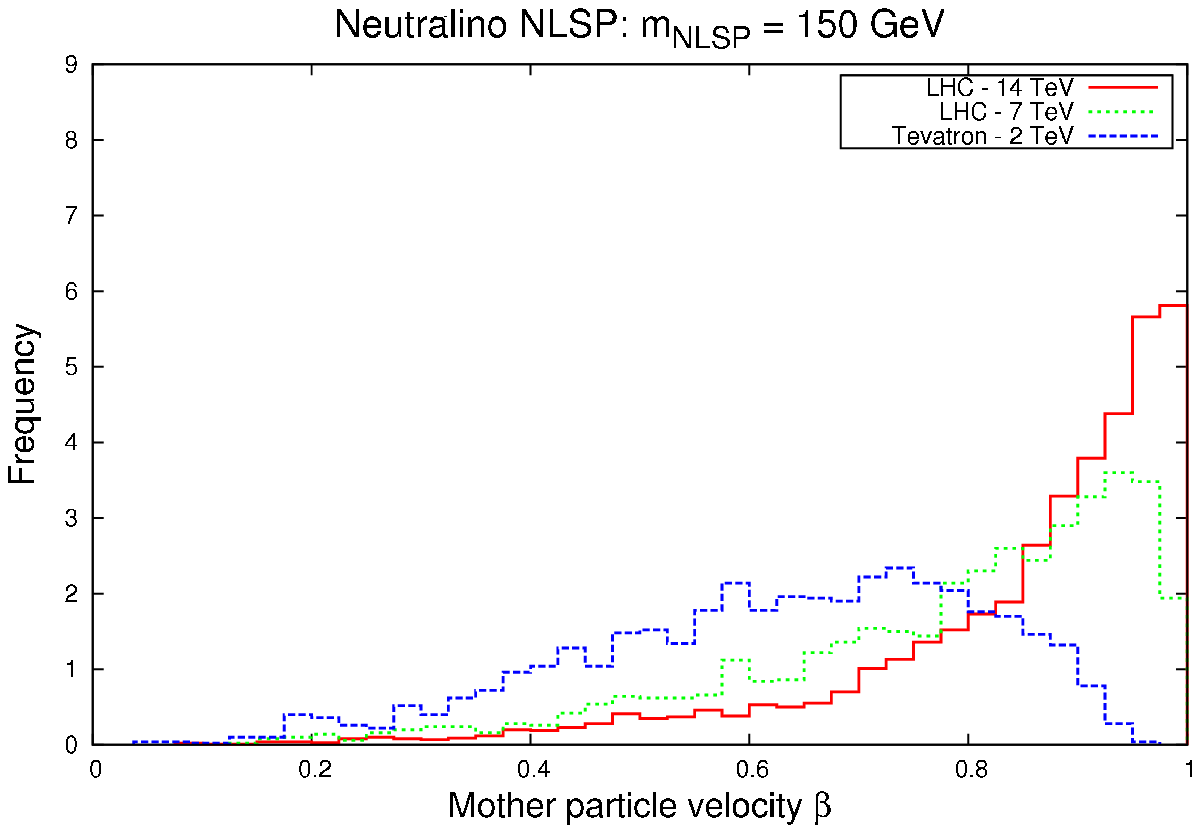} &
\includegraphics[width=\columnwidth]{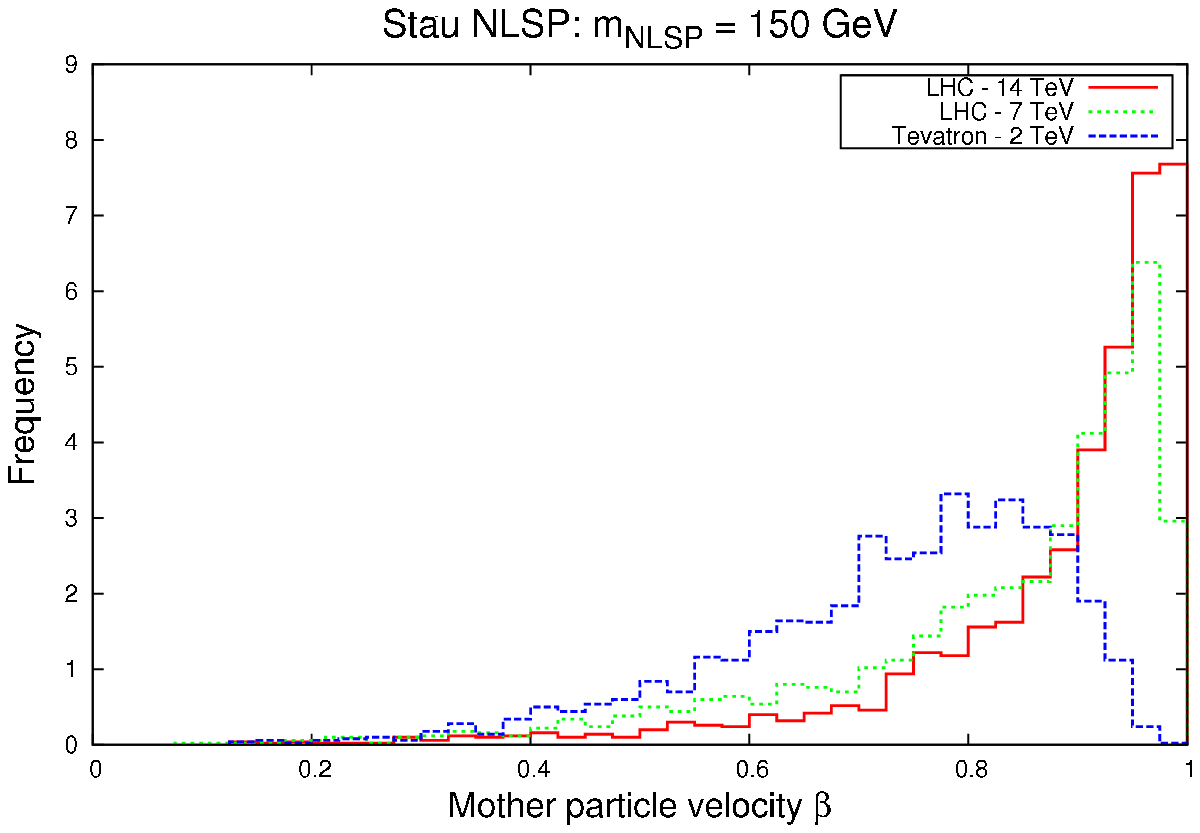}\\
\includegraphics[width=\columnwidth]{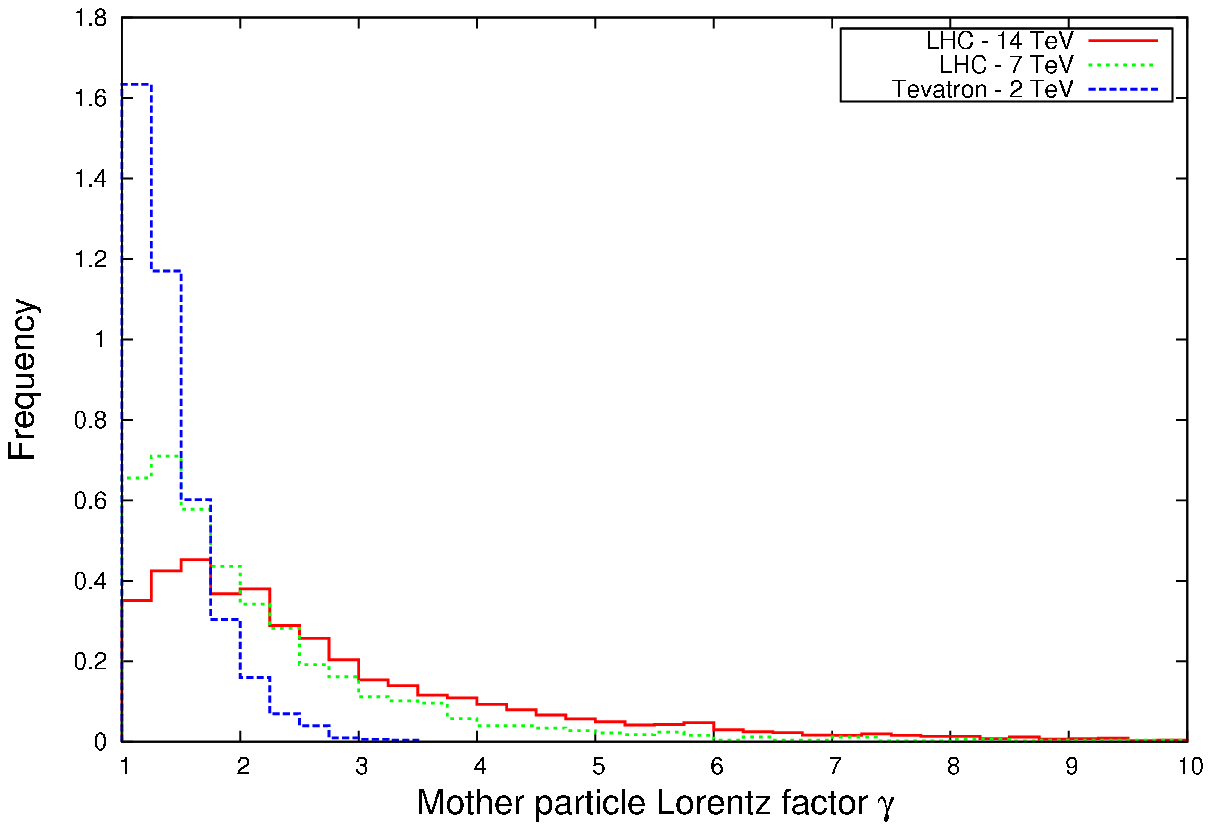} &
\includegraphics[width=\columnwidth]{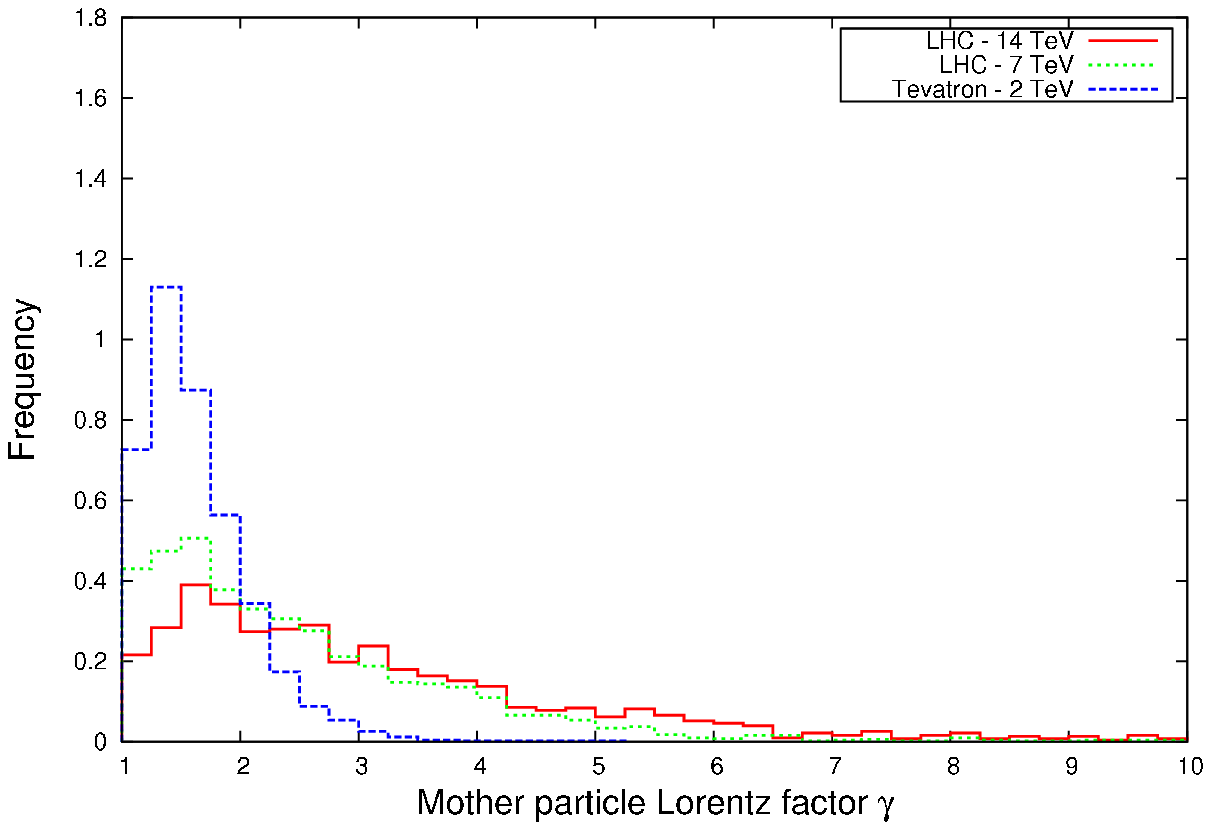}\\
\end{array}$
\caption{Simulated probability distribution functions for the speed
$\beta$ (top panels) and Lorentz factor $\gamma$ (bottom panels) of
mother particles decaying to gravitinos, plotted for the three
collider energies of interest, in the neutralino NLSP (left) and stau
NLSP (right) scenarios.  We have chosen models with ${\mNLSP =
150~\GeV}$ for both NLSP scenarios (this was accomplished by choosing
${\Lambda = 115~\TeV}$ in the neutralino-NLSP scenario and ${\Lambda =
40~\TeV}$ in the stau-NLSP scenario, as well as setting
${\Mmes/\Lambda = 10^3}$ in both scenarios).  The area under each
curve has been normalized to unity.  As expected, increasing the
center-of-mass energy results in the production of faster mother
particles.  Note also that even though $\mNLSP$ is identical for both
of these scenarios, the staus are produced with slightly higher
speeds.  This is because the squark masses (which increase with
increasing $\Lambda$) in the stau-NLSP scenario happen to be slightly
lighter than those in the neutralino-NLSP scenario, for these choices
of GMSB parameters.}
\label{fig:betagamma}
\end{figure*}

\section{Conclusions} 
\label{sec:conclusion}

Light gravitinos in the mass range $\eV$ to $\MeV$ appear in GMSB
models that naturally avoid flavor violation.  We have examined
the decay of supersymmetric particles to light gravitinos at colliders
such as the Tevatron and the LHC.  These decays will give rise to
dramatic signatures, such as prompt di-photons or non-prompt photons,
if the NLSP is a neutralino, or kinked charged tracks or heavy
metastable charged particles, if the NLSP is a stau (or some other
charged particle).  We find large regions of the
gravitino-mass--NLSP-mass parameter space in which the rate for such
events may be appreciable at the Tevatron and LHC and which are
consistent with current null supersymmetry searches.

Given that ${\mG\ll \mNLSP}$ for these events, the
decay kinematics of individual events cannot be used to determine the
gravitino mass.  However, the event rate and the distribution of decay
locations may be used to narrow the range of NLSP and gravitino
masses.  Information about the nature of the NLSP may also be gleaned
from the Standard Model decay products.

One of the attractions of supersymmetry has been its ability to
provide a natural candidate for the cold dark matter required by a
wealth of cosmological observations.  Unfortunately, despite being
well-motivated in GMSB models, the canonical light-gravitino scenario
does not provide a natural cold-dark-matter candidate.  Nevertheless,
this canonical scenario does allow gravitinos with masses ${\mG \alt
30~\eV}$ that compose a fraction of the total dark matter, as
determined by current astrophysical constraints on the relic abundance
and small-scale structure.  Given that upcoming structure formation
observations are expected to probe hot-dark-matter masses as low as
${\mG \sim \eV}$, detection of a gravitino in the mass range ${\eV
\alt \mG \alt 30~\eV}$ via prompt signals at colliders would have
implications for future small-scale-structure measurements.  And
although masses ${\mG \agt 30~\eV}$ are disfavored, they may still be
possible if the pre-BBN history was different than in the canonical
scenario.  Detection of gravitinos in this mass range via non-prompt
and metastable signals at colliders would thus have serious
implications for early-Universe cosmology, and may provide some insight
into the reheating and inflationary eras.  And who knows?
There may indeed be new early-Universe physics that results in a gravitino 
that has the right cosmological abundance and temperature to be the dark
matter. \\

\begin{acknowledgments}
SKL thanks Maria Spiropulu and Sezen Sekmen for their assistance in
setting up the collider simulations.  SKL and MK also thank Mark Wise
for discussion of issues involving the gravitino production rate at
low reheating temperatures.  JLF is grateful to William Molzon and
Daniel Whiteson for helpful conversations.  This work was supported at
Caltech by DoE DE--FG03--92--ER40701 and the Gordon and Betty Moore
Foundation.  The work of JLF was supported in part by NSF grant
PHY--0653656.
\end{acknowledgments}

\end{document}